\documentclass[prb,twocolumn,showpacs]{revtex4}
\usepackage{graphicx}
\usepackage{amsfonts}
\usepackage{amssymb}

\newcommand{\ba}{\begin{eqnarray}}
\newcommand{\be}{\begin{equation}}
\newcommand{\ea}{\end{eqnarray}}
\newcommand{\ee}{\end{equation}}

\newcommand{\trg}{\mathop{\mathrm{trg}}}

\newcommand{\ignore}[1]{}

\begin{document}

\title{Transport through a quantum spin Hall quantum dot}

\author{Carsten Timm}
\email{carsten.timm@tu-dresden.de}
\affiliation{Institute of Theoretical Physics, Technische Universit\"at
Dresden, 01062 Dresden, Germany}

\date{\today}

\begin{abstract}
Quantum spin Hall insulators, recently realized in HgTe/(Hg,Cd)Te quantum
wells, support topologically protected, linearly dispersing edge states with
spin-momentum locking. A local magnetic exchange field can open a gap for the
edge states. A quantum-dot structure consisting of two such magnetic tunneling
barriers is proposed and the charge transport through this device is analyzed.
The effects of a finite bias voltage beyond linear response, of a gate
voltage, and of
the charging energy in the quantum dot are studied within a combination of
Green-function and master-equation approaches. Among other results, a partial
recurrence of non-interacting behavior is found for strong interactions, and
the possibility of controlling the edge magnetization by a locally applied gate
voltage is proposed.
\end{abstract}


\maketitle

\section{Introduction}
\label{sec.intro}

Topological insulators and superconductors\cite{HaK10,RSF10,QiZ11} have recently
become the topic of extensive experimental and theoretical research. These
materials can be described in terms of weakly interacting quasiparticles and
their single-particle Hamiltonians show non-trivial topological properties in
momentum space. They have an energy gap in the bulk but
a topologically protected gapless spectrum of
boundary states. Schnyder \textit{et al.}\cite{SRF08,RSF10} and
Kitaev\cite{Kit09} have put forward an exhaustive classification of these
systems in terms of their Altland-Zirnbauer symmetry classes\cite{Zir96}
and of the number of spatial dimensions.

A topologically nontrivial state is possible for the symplectic class AII
in two dimensions, corresponding to systems with spin-orbit coupling in
the absense of an applied magnetic field and of superconductivity. This
so-called quantum spin Hall (QSH) state\cite{MNZ03,SCN04} has
protected edge states with gapless Weyl-type dispersion. Bernevig \textit{et
al.}\cite{BHZ06} have predicted and K\"onig \textit{et al.}\cite{KWB07,KBM08}
have observed the QSH effect in HgTe/(Hg,Cd)Te quantum wells.

The edge states of the QSH system show spin-mo\-men\-tum locking in the sense
that right-moving (left-moving) electrons are strictly spin-up
(spin-down).\cite{HaK10,QiZ11,KBM08} Consequently, density-density
interactions cannot lead to backscattering since they cannot flip the
spin. On the other hand, spin-dependent scattering, for example by magnetic
impurities, can lead to backscattering.
Unconventional transport properties are thus
expected and possible applications in spintronics can be envisaged. It is
therefore of interest to study electronic transport in prototypical device
geometries involving QSH edge states, which is the subject of this paper.

\begin{figure}
\includegraphics[width=2.6in]{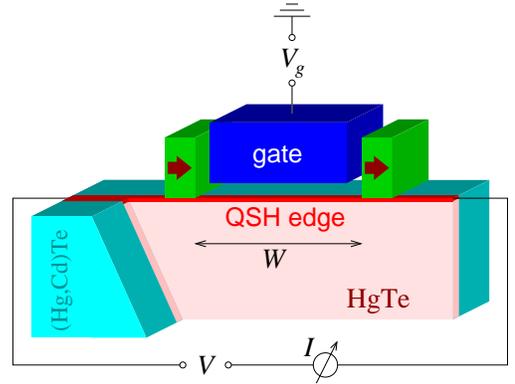}
\caption{(Color online) Cut-away view of the QSH quantum dot.}
\label{fig.sketch}
\end{figure}

The underlying idea is to realize a quantum dot as a finite-length
segment of a QSH edge. This cannot be achieved by electrostatic gating since an
electric potential just shifts the edge bands up or down without opening
a gap and thus does not lead to the formation of tunneling barriers. However,
such barriers could be realized by ferromagnetic insulators grown in contact
with the edge. They would impose an exchange magnetic field orthogonal to the
spin-orbit field and open a
gap. We also include a gate electrode that can be used
to tune the electrostatic potential on the quantum dot. Figure
\ref{fig.sketch} shows a sketch of the device. Since
the QSH edge states typically have a small Fermi wavenumber $k_F$, we consider
the case that the barrier width is negligible compared to $\lambda_F=2\pi/k_F$.
We restrict ourselves to the cases of parallel and antiparallel exchange fields
in the two barriers. Finally, we assume the thermal energy $k_BT$ to be small
compared to the lifetime broadening of the dot levels so that we can set the
temperature to zero. Including a finite temperature is straightforward.
An alternative realization of a QSH quantum dot as the edge of a small QSH
puddle has been analyzed by Tkachov and Hankiewicz,\cite{TkH11} for negligible
charging energy. They have also studied contributions to the resistance due to
the coupling of a QSH edge to normal leads and the effect of an orbital
magnetic field parallel to the spin-orbit field.\cite{TkH11}

After introducing the model in Sec.\ \ref{sec.model}, we study the effect
of the gate voltage and the bias voltage on the transport in Sec.\
\ref{sec.trans1}. A Landauer approach\cite{Lan57} is
used to obtain the current for arbitrary strength of the magnetic barriers. We
will see that the result can also be analyzed in terms
of a generalized Meir-Wingreen (MW) formula for the current\cite{MeW92,JWM94} in
terms of the non-interacting Green function. It is generalized in the sense of
pertaining to a real-space continuum Hamiltonian instead of a tunneling
Hamiltonian, which leads to a nontrivial form of the dot-lead coupling
functions $\Gamma^L$, $\Gamma^R$.

In the next step, the effect of electron-electron interaction on the dot is
studied in Sec.\ \ref{sec.trans2}. We again employ a simple model by including a
charging energy in terms of the excess charge. This is
valid if the range of the electron-electron interaction is
large compared to the dot size $W$, i.e., for small dots, but should give
qualitatively correct results beyond this regime.
The opposite case of short-range interactions is certainly of interest. For an
unmodulated QSH edge this case has
been studied for example in Refs.\ \onlinecite{WBZ06}, \onlinecite{XuM06},
\onlinecite{BDR11}. A quantum dot in a spinless Luttinger liquid, not a QSH
edge, has been studied by several
authors.\cite{KaF92,FuN93,ChW93,MEA05} Tunneling through a quantum dot between
\emph{two} QSH edges described as Luttinger liquids has been addressed by Law
\textit{et al.}\cite{LSL10}

In Sec.\ \ref{sec.trans2}, we employ an equation-of-motion approach to
obtain the multiple-pole structure of the spectral function appearing in the
generalized MW formula. We discuss possible approximations for the
weights of the poles and show results calculated from a master equation in the
sequential-tunneling approximation with additional level broadening, valid for
weak coupling through the magnetic barriers. We summarize the main results in
Sec.~\ref{sec.sum}.

\section{Model}
\label{sec.model}

The non-interacting part of the Hamiltonian of the QSH edge with magnetic
barriers is written as a $2\times 2$ matrix in spin space,
\ba
H_0 & = & -i \hbar v_F \sigma^z \partial_x
  - \eta \hbar v_F \sigma^x \delta(x)
  \mp \eta \hbar v_F \sigma^x \delta(x-W) \nonumber \\
&& {}+ V(x) ,
\label{2.H0}
\ea
where $v_F$ is the Fermi velocity, $\sigma^x$ and $\sigma^z$ are Pauli
matrices, $\eta$ is the dimensionless strength of the magnetic barriers,
and $V(x)$ is a non-uniform electric potential. A $2\times 2$ unit
matrix is implied in the last term. $H_0$ is certainly only
valid within the bulk energy gap. The first term is the Weyl Hamiltonian of the
bare edge.\cite{KBM08,VaO11} We neglect
higher-order spatial derivatives, which lead to non-linear terms in the
dispersion. An electric field perpendicular to the layers induces an additional
Rashba spin-orbit-coupling term $H_R = -i \alpha \sigma^y \partial_x$.
However, this term can be absorbed into the Weyl term by a rotation in spin
space.\cite{VaO11} The upper (lower) sign of the third term refers to parallel
(antiparallel) exchange fields in the two barriers. The parameter $\eta$ can
be written as
\be
\eta = \frac{g\mu_B B_\mathrm{exc} L}{\hbar v_F} ,
\ee
where $g$ is the g-factor, $\mu_B$ is the Bohr magneton, $B_\mathrm{exc}$ is
the exchange field, and $L$ is the width of the magnetic barrier. Equation
(\ref{2.H0}) is obtained in the limit $B_\mathrm{exc}\to\infty$, $L\to 0$,
keeping $B_\mathrm{exc}L$ constant. For the potential $V(x)$ we take
\be
V(x) = \left\{\begin{array}{ll}
  eV/2 & \mbox{for $x\le 0$} , \\[0.5ex]
  -eV_g & \mbox{for $0 < x \le W$} , \\[0.5ex]
  -eV/2 & \mbox{for $x > W$} ,
  \end{array}\right.
\ee
where $V$ is the bias voltage and $V_g$ is the gate voltage.

Since the time-independent Schr\"odinger equation resulting from $H_0$ is of
first order in spatial derivatives, we can solve it by means of a (non-unitary)
``spatial-evolution operator.'' Multiplying the Schr\"odinger equation by
$\sigma^z$ from the left, we obtain
\ba
\partial_x \psi
& = & i\,\frac{E-V(x)}{\hbar v_F}\, \sigma^z\, \psi(x) \nonumber \\
&& {}- \eta\, \sigma^y\, \delta(x)\, \psi(x)
  \mp \eta\, \sigma^y\, \delta(x-W)\, \psi(x) ,
\ea
which for $x\ge x_0$ is solved by
\ba
\psi(x) & = & S_\leftarrow\, \exp\bigg( \int_{x_0}^x dx'\, \bigg[
  i\, \frac{E-V(x')}{\hbar v_F}\, \sigma^z \nonumber \\
&& {}- \eta\, \sigma^y\, \delta(x')
  \mp \eta\, \sigma^y\, \delta(x'-W) \bigg]\bigg)\, \psi(x_0) ,
\label{2.Uspace}
\ea
where $S_\leftarrow$ is a spatial-ordering directive; operators acting on
$\psi(x_0)$ are ordered with their spatial coordinates increasing from right to
left. We can obtain the boundary condition at the barrier at $x=0$ by making
$x_0$ infinitesimally negative and $x$ infinitesimally positive,
\be
\psi(0^+) = e^{-\eta \sigma^y}\, \psi(0^-)
  = (\cosh \eta - \sigma^y \sinh \eta)\, \psi(0^-) .
\label{2.BC1}
\ee
Since the Schr\"odinger equation is of first order in spatial derivatives, there
is only a single boundary condition.\cite{footnote.BC} Analogously, we find the
boundary condition at the other barrier, $\psi(W^+) = (\cosh \eta \mp \sigma^y
\sinh \eta)\, \psi(W^-)$.

To study transport, we assume the states in the leads to be filled up to the
chemical potential $\mu$, measured relative to the Weyl nodes (band crossing
points), which are shifted by the potential $\pm eV/2$. We take $\mu$ to be
the same in the two leads since they are parts of the edge of the same quantum
well.

In the following, we will need the eigenstates of the decoupled dot, which
corresponds to the limit $\eta\to\infty$. The solution is straightforward: The
Schr\"odinger equation for the eigenspinors $\psi_\nu(x)$ to eigenenergies
$E_\nu$ reads
\be
-i \hbar v_F \sigma^z\, \partial_x \psi_\nu
  - eV_g\, \psi_\nu(x) = E_\nu\, \psi_\nu(x)
\ee
with the boundary conditions
\be
(1+\sigma^y)\, \psi_\nu(0) = 0 , \qquad
(1\mp\sigma^y)\, \psi_\nu(W) = 0 ,
\ee
which give
\be
\psi_{\nu\downarrow}(0) = -i\, \psi_{\nu\uparrow}(0), \qquad
\psi_{\nu\downarrow}(W) = \pm i\, \psi_{\nu\uparrow}(W) .
\ee
The normalized solutions are
\be
\psi_\nu(x) = \frac{1}{\sqrt{2W}} \left(\begin{array}{c}
  \exp\left(i\, \frac{E_\nu + eV_g}{\hbar v_F}\, x\right) \\[1.5ex]
  -i\, \exp\left(-i\, \frac{E_\nu + eV_g}{\hbar v_F}\, x\right)
  \end{array}\right)
\ee
with
\ba
\lefteqn{ E_\nu = -eV_g + \frac{\pi\hbar v_F}{W} } \nonumber \\
&& {}\times \left\{\begin{array}{ll}
    \nu + 1/2 & \mbox{for parallel exchange fields}, \\
    \nu & \mbox{for antiparallel exchange fields},
  \end{array}\right.
\label{2.Enu}
\ea
where $\nu$ can assume any integer value. Note that the spectrum is an
equidistant ladder, which is, unlike for the harmonic oscillator, unbounded
from both above and below. The level spacing is $E_0 := \pi\hbar v_F/W$. Of
course, the discrete spectrum only exists inside the bulk gap and is only
equidistant as long as non-linear terms in the dispersion of the edge states can
be neglected. In the absense of a gate voltage, the spectrum is symmetric with
respect to zero energy for both orientations of the exchange fields due to the
particle-hole symmetry of $H_0$. For antiparallel exchange fields, one
eigenstate has energy zero.

Finally, the Coulomb interaction is described by the particle-hole-symmetric
term $H_\mathrm{int} =({e^2}/{2C})\, (\Delta n)^2$,
where $C$ is the capacitance of the quantum dot and $\Delta n$ is the excess
number of electrons on the dot compared to neutrality. In terms of the number
operators $n_\nu = c_\nu^\dagger c_\nu$ of single-particle dot states
$|\nu\rangle$, we have
\be
H_\mathrm{int} = \frac{e^2}{2C}\, \bigg[ \sum_\nu \bigg(n_\nu - \frac12\bigg)
  \bigg]^2 .
\label{2.Hint}
\ee

\section{Transport through a non-interacting dot}
\label{sec.trans1}

Neglecting the electron-electron interaction, the current through the QSH
quantum dot can be expressed in terms of its transmission coefficient $T(E)$ by
the Landauer formula,\cite{Lan57}
\be
I = \frac{e}{h}\, \int_{\mu-eV/2}^{\mu+eV/2} dE\: T(E) .
\label{3.Landauer}
\ee
Since $T(E)$ does no depend on the bias voltage $V$ in our case, the
differential conductance is simply given by
\be
\frac{dI}{dV} = \frac{e^2}{h}\, \frac{T(\mu+eV/2) + T(\mu-eV/2)}{2} .
\label{3.dIdV}
\ee
The transmission coefficient $T(E)$ is obtained from the transfer matrix
$\mathcal{T}=\mathcal{T}_R \mathcal{T}_\mathrm{dot} \mathcal{T}_L$ of the
device, where the three factors are the transfer matrices of the right barrier,
the dot region, and the left barrier, respectively. Since the
right-moving (left-moving) electrons have spin up (down), Eq.\ (\ref{2.BC1})
implies that the transfer matrix for the left barrier is
\be
\mathcal{T}_L = \cosh \eta - \sigma^y \sinh \eta
  = \left(\begin{array}{cc}
      \cosh \eta & i\, \sinh \eta \\
      -i\, \sinh \eta & \cosh \eta
    \end{array}\right) .
\ee
(The transmission coefficient of a single barrier thus equals $1/\cosh^2\eta$.)
For the right barrier we get
\be
\mathcal{T}_R = \left(\begin{array}{cc}
      \cosh \eta & \pm i\, \sinh \eta \\
      \mp i\, \sinh \eta & \cosh \eta
    \end{array}\right) .
\ee
The transfer matrix for the dot region can be infered from the
spatial-evolution operator in Eq.\ (\ref{2.Uspace}) for $x_0=0^+$, $x=W^-$,
\ba
\lefteqn{ \mathcal{T}_\mathrm{dot} = \exp\left( i\, \frac{E+eV_g}{\hbar
  v_F}\, \sigma^z W \right) } \nonumber \\
&& = \left(\begin{array}{cc}
    \exp\left(i\, \frac{E+eV_g}{\hbar v_F}\, W\right) & 0 \\
    0 & \exp\left(-i\, \frac{E+eV_g}{\hbar v_F}\, W\right)
    \end{array}\right) .\quad
\ea
The transmission coefficient $T = |t|^2$ is obtained in the standard way by
solving
\be
\mathcal{T}_R \mathcal{T}_\mathrm{dot} \mathcal{T}_L\,
  \left(\begin{array}{c}
    1 \\ r
  \end{array}\right)
  = \left(\begin{array}{c}
    t \\ 0
  \end{array}\right)
\ee
for $t$. The result is
\be
T(E) = \frac{1}{1 + \trg^2 \left(\frac{E+eV_g}{\hbar v_F}\, W\right) \,
  \sinh^2 2\eta } ,
\label{3.T}
\ee
where $\trg = \cos$ ($\sin$) for parallel (antiparallel) exchange
fields. The transmission is maximal whenever the energy $E$ coincides with an
eigenenergy $E_\nu$ of the decoupled dot. In this situation, the transmission
coefficient is unity so that the device is perfectly transparent on resonance.
The width of the maxima is controlled by the strength of the magnetic barriers,
$\eta$, with the width getting exponentially small for large $\eta$. This
is more easily seen by writing the transmission coefficient as a series,
\be
T(E) = \frac{E_0}{2\pi\,\cosh 2\eta} \sum_{\nu=-\infty}^\infty
  \frac{2\gamma}{(E-E_\nu)^2+\gamma^2} ,
\label{3.Tseries}
\ee
where the width of the Lorentzian peaks is
\be
\gamma := \frac{E_0}{\pi}\, \ln\coth \eta .
\ee
An analytical expression for the current can be obtained by inserting
$T(E)$ into Eq.\ (\ref{3.Landauer}). We defer this to the next section.

\begin{figure}
\raisebox{3ex}{(a)}\includegraphics[width=2.6in]{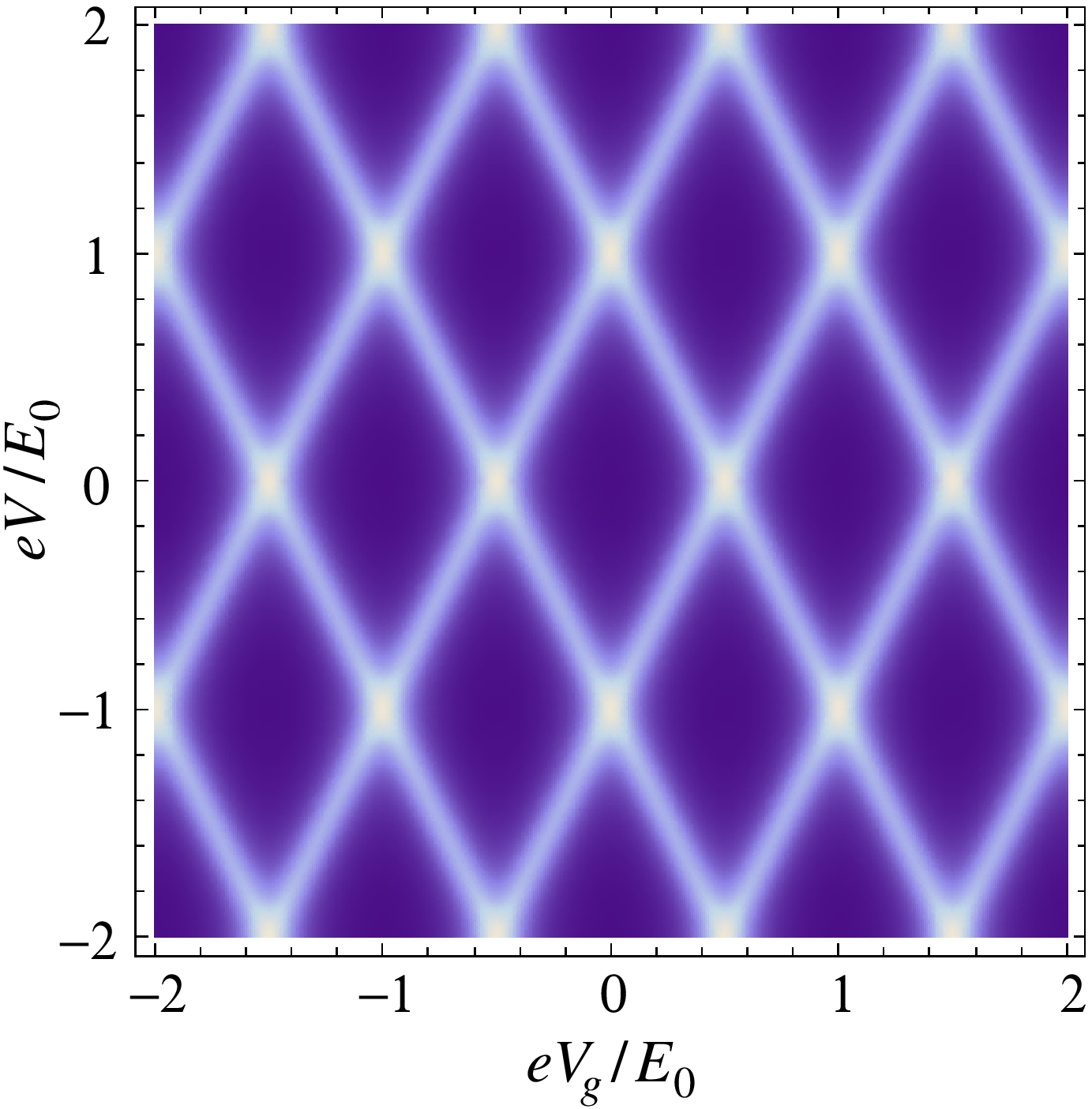}
\raisebox{3ex}{(b)}\includegraphics[width=2.6in]{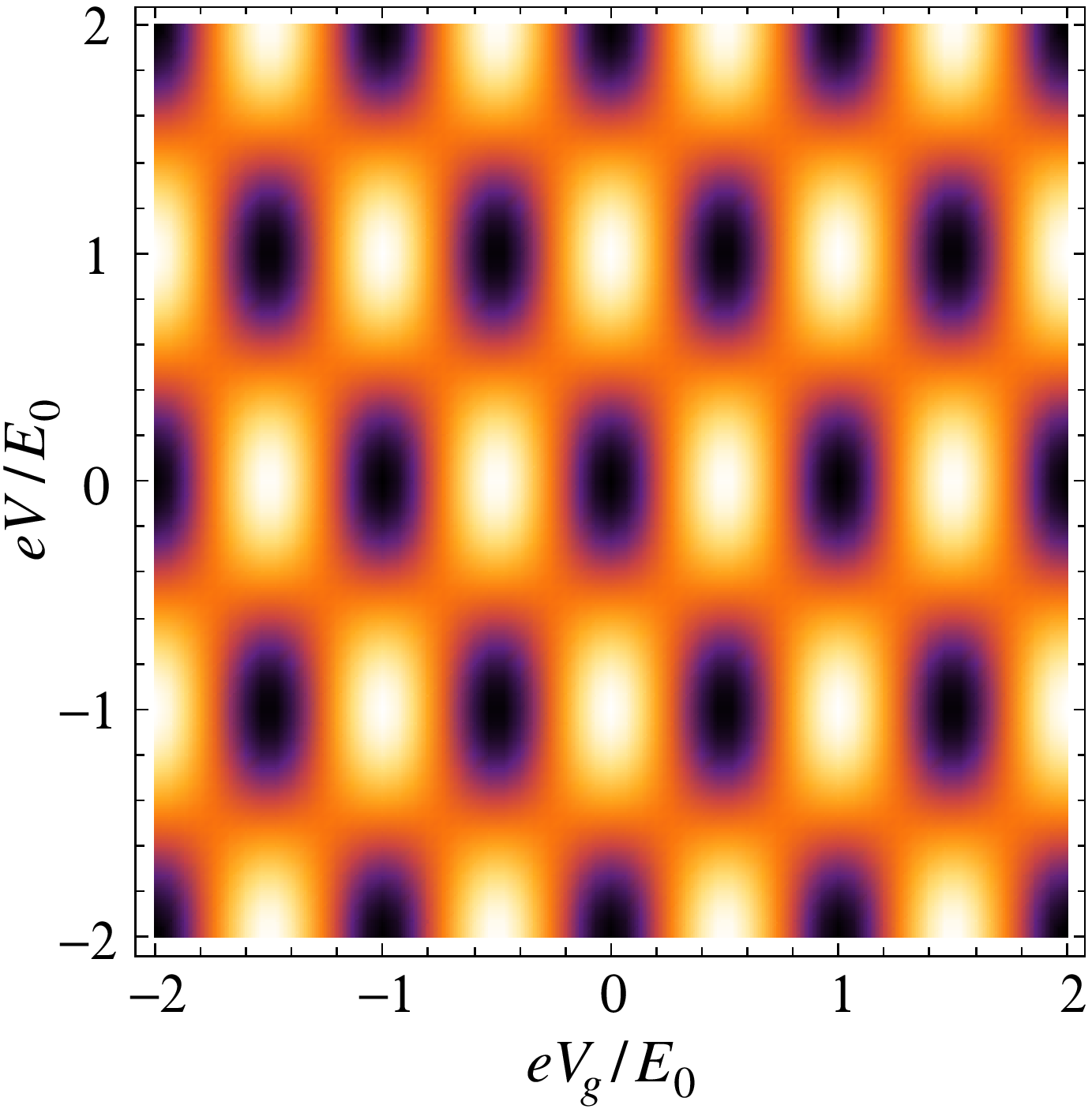}
\raisebox{2ex}{(c)}\includegraphics[width=3.0in,clip]{QSHQD_fig2c.eps}
\caption{(Color online) Density plots of the differential conductance
$dI/dV$ through a QSH quantum dot with parallel exchange fields as a function of
gate voltage $V_g$ and bias voltage $V$ in the absence of electron-electron
interaction. The dimensionless strength of the magnetic barriers was set to
(a) $\eta=1$ and (b) $\eta=0.1$, respectively. Light colors denote large
$dI/dV$. (c) Linear conductance $G = dI/dV|_{V=0}$ as a function of the gate
voltage $V_g$. The level spacing
of the decoupled dot, $E_0 = \pi\hbar v_F/W$, is used as the energy unit.}
\label{fig.noninter}
\end{figure}

This is not required for deriving the differential
conductance $dI/dV$, which can immediately be read off from Eqs.\
(\ref{3.dIdV}) and (\ref{3.T}). Note that the chemical potential $\mu$ and
the gate voltage $V_g$ only appear in the combination $eV_g+\mu$ since only
energy \emph{differences} between the dot and the leads enter. We can thus set
$\mu=0$ without loss of generality. $dI/dV$ is plotted in Fig.\
\ref{fig.noninter} as a function of gate and bias voltages. The figure clearly
exhibits the suppression of the low-bias conductance off resonance due to Pauli
blockade. The Pauli-blockade diamonds are filled in for strong coupling
to the leads (weak magnetic barriers), and for $\eta\to 0$ we recover the
constant conductance $dI/dV = e^2/h$ of an open channel. $dI/dV$ is periodic
in the gate voltage with period $E_0=\pi\hbar v_F/W$ and in the bias voltage
with period $2E_0$. Also, going from parallel to antiparallel exchange fields
has the same effect as shifting $eV_g$ by half a period, $E_0/2$.

Additional insight that will also prove useful for the interacting case can be
gained by writing the current in the form of the MW formula\cite{MeW92}
\be
I = \frac{e}{h} \int_{\mu-eV/2}^{\mu+eV/2} dE\: \mathrm{Im}\, \mathrm{Tr}\,
  \hat\Gamma\, (-2)\, \hat{G}^\mathrm{ret}_0(E) ,
\label{3.I_MWgeneral}
\ee
where
\be
\hat\Gamma = \hat\Gamma^L (\hat\Gamma^L+\hat\Gamma^R)^{-1} \hat\Gamma^R ,
\ee
$\hat\Gamma^L$ and $\hat\Gamma^R$ are matrices of coupling functions to the
left and right leads, and $\hat{G}^\mathrm{ret}_0(E)$ is a matrix of retarded
non-interacting Green functions. This form is general under the condition that
$\hat\Gamma^L$ and $\hat\Gamma^R$ differ at most by a scalar factor.\cite{MeW92}
We will use the basis of decoupled dot states $|\nu\rangle$. Equation
(\ref{3.Landauer}) can be written in this form by taking
\ba
\Gamma^\alpha_{\nu\nu'} & = & \delta_{\nu\nu'}\,\Gamma^\alpha , \\
G^\mathrm{ret,0}_{\nu\nu'}(E) & = & \frac{\delta_{\nu\nu'}}
  {E - E_\nu - \Sigma^\mathrm{ret}_0} ,
\ea
where the self-energy due to the coupling to the leads is independent of $\nu$.
Hence, we can write
\be
I = \frac{e}{h} \int_{\mu-eV/2}^{\mu+eV/2} dE\: \Gamma
  \sum_\nu (-2)\, \mathrm{Im}\,
  \frac{1}{E - E_\nu - \Sigma^\mathrm{ret}_0}
\label{3.I_MW}
\ee
with $\Gamma = \Gamma^L\Gamma^R/(\Gamma^L+\Gamma^R)$, which for our case of
symmetric coupling implies $\Gamma=\Gamma^L/2=\Gamma^R/2$. Comparison with
Eqs.\ (\ref{3.Landauer}) and (\ref{3.Tseries}) yields
\ba
\Gamma & = & \frac{\hbar v_F}{2W}\, \frac{1}{\cosh 2\eta} , \\
\Sigma^\mathrm{ret}_0 & = & -i\,\gamma \; = \;
 -i\, \frac{\hbar v_F}{W}\, \ln\coth \eta .
\ea
We note in passing that the same decay constant $\gamma$ is found by solving
the time-dependent Schr\"odinger equation for the \emph{coupled} quantum dot
under boundary conditions implying that no probability flows into the dot.

In the standard MW formula the non-interacting self-energy is given
by $-i(\Gamma^L+\Gamma^R)/2=-2i\Gamma$.\cite{JWM94} Our results contains a
crucial difference: We have
\be
\Sigma^\mathrm{ret}_0
  = -i\, \frac{\hbar v_F}{2W}\, \ln \frac{\frac{\hbar v_F}{2W} + \Gamma}
  {\frac{\hbar v_F}{2W} - \Gamma} ,
\label{3.Sigma_Gamma}
\ee
which does not coincide with $-2i\Gamma$. This is
because the MW formula was derived for a tunneling Hamiltonian,\cite{MeW92}
whereas our Hamiltonian
$H_0$ is not of tunneling form. Our model should approach a tunneling
Hamiltonian in the limit of weak coupling, i.e., of strong magnetic barriers,
$\eta\gg 1$. In this limit we obtain
\be
\Gamma \cong \frac{\hbar v_F}{W}\, e^{-2\eta} , \qquad
\Sigma^\mathrm{ret}_0 \cong -2i\, \frac{\hbar v_F}{W}\, e^{-2\eta}
\ee
so that we indeed recover the standard relation between the self-energy and the
coupling to the leads.

It is remarkable that the current can be given in a generalized MW
form, Eq.\ (\ref{3.I_MW}), even when the description in terms of a tunneling
Hamiltonian is not valid. The difference between a real-space first-quantized
Hamiltonian and the corresponding approximate tunneling Hamiltonian has been
discussed by Appelbaum\cite{App67} and more recently by Patton.\cite{Pat10} Both
authors point out that the current calculated from the approximate tunneling
Hamiltonian is only correct to second order in the tunneling amplitudes, i.e.,
to first order in $\Gamma$.\cite{App67,Pat10} We have here obtained a MW-type
current formula for a real-space Hamiltonian that is exact for any coupling, at
least for non-interacting electrons.
We can also turn the argument around: The QSH quantum dot can be described by a
tunneling Hamiltonian with dot-lead couplings $\Gamma^\alpha$ \emph{if} we use
the renormalized self-energy, Eq.~(\ref{3.Sigma_Gamma}).

\section{Transport through an interacting dot}
\label{sec.trans2}

In the present section we consider the effect of electron-electron interaction
in the form of a charging energy $H_\mathrm{int}$, see Eq.\ (\ref{2.Hint}).
Since this term preserves particle-hole symmetry, the current is still odd and
the differential conductance is even in the bias voltage.

To obtain the current for the interacting case, we replace the non-interacting
Green function $\hat G^\mathrm{ret}_0$ in the generalized MW formula by the
interacting one, $\hat G^\mathrm{ret}$. In principle, we should start from the
general expression (\ref{3.I_MWgeneral}), which allows for the Green-function
matrix being non-diagonal in the dot single-particle basis $\{|\nu\rangle\}$. We
will argue below that it remains diagonal in the presence of $H_\mathrm{int}$. A
simple argument for this is that $H_\mathrm{int}$ does not induce any scattering
from one state $|\nu\rangle$ into another and thus leaves the self-energy
diagonal in $\nu$. In this case, we can replace Eq.\ (\ref{3.I_MW}) by
\be
I = \frac{e}{h} \int_{\mu-eV/2}^{\mu+eV/2} dE\: \Gamma
  \sum_\nu (-2)\, \mathrm{Im}\, G^\mathrm{ret}_{\nu\nu}(E) .
\label{4.I_MW}
\ee
The simplest nontrivial approximation for $\hat G^\mathrm{ret}$
involves a Hartree-Fock decoupling of the interaction term. In this
approximation, the spectral function $A_\nu(E) = -2\,\mathrm{Im}\,
G^\mathrm{ret}_{\nu\nu}(E)$ still is a Lorentzian at a shifted energy,
which includes the \emph{average} Coulomb interaction with the other electrons
on the dot.\cite{footnote.HF} This can only be valid if the spectral function
is dominated by a single peak. We now show that this is generally not the case.

To go beyond the Hartree-Fock approximation, we employ an equation-of-motion
approach. The crucial assumption is that the effect of the
coupling to the leads is not changed by the interaction. This is motivated by
the fact that all dot single-particle states couple equally to the leads
and that Slater determinants of these states remain eigenstates of the
interacting dot. We implement this approximation by using a non-hermitian
dot Hamiltonian
\be
K_\mathrm{dot} = \sum_\nu (E_\nu - i\,\gamma)\,  n_\nu + H_\mathrm{int} .
\label{4.K}
\ee
The equation of motion for the full Green function is then
\ba
\lefteqn{ (E-E_\nu+i\,\gamma)\, G^\mathrm{ret}_{\nu\nu'}(E) = \delta_{\nu\nu'}
  + \frac{i}{\hbar} \int_0^\infty dt\, e^{iEt/\hbar} } \nonumber \\
&& {}\times \left\langle \big\{ e^{iK_\mathrm{dot}t/\hbar}\,
  [H_\mathrm{int}, c_\nu] \, e^{-iK_\mathrm{dot}t/\hbar} ,
  c_{\nu'}^\dagger \big\} \right\rangle , \qquad
\label{4.EOM}
\ea
where $c_\nu^\dagger$ ($c_\nu$) creates (annihilates) an electron in dot state
$|\nu\rangle$. The solution of Eq.\ (\ref{4.EOM}) is relegated to Appendix
\ref{app.eom}. The result for the diagonal components of the Green function is
\ba
G^\mathrm{ret}_{\nu\nu}(E) & = &
  \sum_{\ldots,m_{\nu-1},m_{\nu+1},\ldots=0}^1 \nonumber \\
&& {}\times \frac{P_\nu(\ldots,m_{\nu-1},m_{\nu+1},\ldots)}
  {E-E_\nu - \frac{e^2}{C} \sum_{\mu\neq\nu}
  \left( m_\mu - \frac{1}{2} \right) + i\gamma} , \qquad
\label{4.Gnunu}
\ea
where the sum is over all $m_\mu$ with $\mu\neq \nu$ and
\ba
\lefteqn{ P_\nu(\ldots,m_{\nu-1},m_{\nu+1},\ldots) } \nonumber \\
&& := \left\langle \prod_{\mu\neq \nu}
  \left\{\begin{array}{cl}
    1-n_\mu & \mbox{for $m_\mu=0$} \\
    n_\mu   & \mbox{for $m_\mu=1$}
  \end{array}\right\} \right\rangle
\label{4.Pnu}
\ea
is the probability of the dot occupation-number state
$|\ldots,m_{\nu-1},m_{\nu+1},\ldots\rangle$ without reference to the
occupation of single-particle state $|\nu\rangle$. Introducing the probabilities
$P(\ldots,m_0,m_1,\ldots)$ of the full dot occupation-number states
$|\ldots,m_0,m_1,\ldots\rangle$, we can write
\ba
\lefteqn{ P_\nu(\ldots,m_{\nu-1},m_{\nu+1},\ldots) } \nonumber \\
&& = P(\ldots,m_{\nu-1},0,m_{\nu+1},\ldots) \nonumber \\
&& {}+ P(\ldots,m_{\nu-1},1,m_{\nu+1},\ldots) .
\label{4.PPP}
\ea
It is also argued in Appendix \ref{app.eom} that the off-diagonal components
$G^\mathrm{ret}_{\nu\nu'}(E)$ vanish in the stationary state since they would
require off-diagonal components of the stationary reduced density operator.

The retarded Green function $G^\mathrm{ret}_{\nu\nu}(E)$, Eq.\ (\ref{4.Gnunu}),
has a multiple-pole structure with one pole for every possible excess electron
number $\Delta n_{\neg\nu} := \sum_{\mu\neq\nu} (m_\mu-1/2)$ in the other
single-particle states $|\mu\rangle \neq |\nu\rangle$.\cite{KuC07}
The energy of each pole contains the interaction energy of an electron in state
$|\nu\rangle$ with the excess electron number $\Delta n_{\neg\nu}$. The weight
of each pole is the probability to find that value of
$\Delta n_{\neg\nu}$. This structure is highly
plausible as the generalization of the two-pole structure for the Anderson
model.\cite{MWL91} In the
non-interacting limit, $e^2/C\to 0$, the peaks merge into a single one at the
single-particle energy. The interacting spectral function
\ba
A_\nu(E) & = & \!\!\!\! \sum_{\ldots,m_{\nu-1},m_{\nu+1},\ldots=0}^1
  \!\!\!\! P_\nu(\ldots,m_{\nu-1},m_{\nu+1},\ldots) \nonumber \\
&& {}\times \frac{2\gamma}
  {\left[E-E_\nu - \frac{e^2}{C} \sum_{\mu\neq\nu}
  \left( m_\mu - \frac{1}{2} \right)\right]^2 + \gamma^2}
\label{4.Anu}
\ea
is a sum of Lorentzians, all of the same width $\gamma$, thanks to our main
approximation.

Inserting the spectral function into the generalized MW formula (\ref{4.I_MW})
we obtain the current
\begin{widetext}
\ba
I & = & \frac{e}{h} \int_{\mu-eV/2}^{\mu+eV/2} dE\: \Gamma
  \sum_\nu A_\nu(E)
\; = \; \frac{e}{h} \, \frac{\hbar v_F}{W} \, \tanh\frac{W\gamma}{\hbar v_F} \,
  \sum_\nu \sum_{\ldots,m_{\nu-1},m_{\nu+1},\ldots=0}^1
  P_\nu(\ldots,m_{\nu-1},m_{\nu+1},\ldots) \nonumber \\
&& {}\times \left( \arctan \frac{E_\nu + \frac{e^2}{C} \sum_{\mu\neq\nu}
    \left( m_\mu - \frac{1}{2} \right) - \mu + \frac{eV}{2}}{\gamma}
  - \arctan \frac{E_\nu + \frac{e^2}{C} \sum_{\mu\neq\nu}
    \left( m_\mu - \frac{1}{2} \right) - \mu - \frac{eV}{2}}{\gamma}
  \right) \nonumber \\
& = & \frac{e}{h} \, \frac{\hbar v_F}{W} \, \tanh\frac{W\gamma}{\hbar v_F} \,
  \sum_\nu \sum_{\ldots,m_{\nu-1},m_{\nu+1},\ldots=0}^1
  P_\nu(\ldots,m_{\nu-1},m_{\nu+1},\ldots) \nonumber \\
&& {}\times
  \left( \arctan \frac{\frac{\pi\hbar v_F}{W}\, \big\{ { \nu+1/2 \atop \nu }
   \big\} - eV_g + \frac{e^2}{C} \sum_{\mu\neq\nu}
    \left( m_\mu - \frac{1}{2} \right) - \mu + \frac{eV}{2}}{\gamma} \right.
  \nonumber \\
&& \quad\left.{}- \arctan \frac{\frac{\pi\hbar v_F}{W}\,
    \big\{ { \nu+1/2 \atop \nu } \big\}
    - eV_g + \frac{e^2}{C} \sum_{\mu\neq\nu}
    \left( m_\mu - \frac{1}{2} \right) - \mu - \frac{eV}{2}}{\gamma}
  \right) ,
\label{4.Ifinal}
\ea
\end{widetext}
where the upper (lower) expression in curly braces correspond to parallel
(antiparallel) exchange fields, see Eq.\ (\ref{2.Enu}). Unlike in the
non-interacting case, the differential conductance $dI/dV$ is not simply given
by the difference of the integrand at $\mu+eV/2$ and $\mu-eV/2$, since the
probabilities $P_\nu$ generally depend on the bias voltage.

The next task is to find the probabilities
$P_\nu$. The simplest approximation would be to neglect
correlations between occupations of single-particle states by writing
\ba
\lefteqn{ P_\nu(\ldots,m_{\nu-1},m_{\nu+1},\ldots) } \nonumber \\
&& \cong \prod_{\mu\neq \nu}
  \left\{\begin{array}{cl}
    \langle 1-n_\mu \rangle & \mbox{for $m_\mu=0$} \\
    \langle n_\mu \rangle   & \mbox{for $m_\mu=1$}
  \end{array}\right\} .
\ea
Unlike the Hartree-Fock approximation, this retains the correct multiple-pole
structure, thus transport features resulting from them appear in the right place
in the conductance plots. However, neglecting correlations due to Coulomb
repulsion overestimates charge fluctuations. This means that side-peaks in the
spectral function carry too much weight.

A better approximation is obtained by realizing that the probabilities
$P$ of many-particle states, which determine the peak weights
$P_\nu$ through Eq.\ (\ref{4.PPP}), are the stationary
solutions of a Pauli master equation, i.e., a master equation for the
diagonal components of the reduced density operator. At least for a
tunneling Hamiltonian, an exact Pauli master equation can in principle be
obtained by eliminating the off-diagonal components.\cite{Zwa60,KLW10,Tim11}
In practice, one has to rely on approximations. For weak coupling to the leads
but arbitrary interaction strengths one can retain
only the terms of leading order in the coupling, which constitutes the
sequential-tunneling approximation.

In the strict sequential-tunneling approximation, the broadening of dot states
due to dot-lead coupling is neglected so that the probabilities show
unphysical jumps as a function of bias voltage. We therefore include
the broadening $\gamma$ beyond sequential tunneling.\cite{ScS94} Denoting the
occupation-number states by $\vec m := |\ldots,m_0,m_1,\ldots\rangle$,
the sequential-tunneling rates $R_{\vec m\to\vec m'}$ vanish unless $\vec m$
and $\vec m'$ differ in exactly one of the occupation numbers $m_\mu$. If
$|\nu\rangle$ is the corresponding single-particle state, the rate is
\begin{widetext}
\ba
R_{\vec m\to\vec m'} & = & R_\nu^\mathrm{in}
\; := \; R_0 \left( \int_{-\infty}^{\mu-eV/2}
  + \int_{-\infty}^{\mu+eV/2} \right) \frac{dE}{2\pi} \,
  \frac{2\gamma}{\big[E-E_\nu - \frac{e^2}{C} \sum_{\mu\neq\nu}
  \left( m_\mu - \frac{1}{2} \right)\big]^2 + \gamma^2} \nonumber \\
& = & \frac{R_0}{\pi} \left( \pi
  - \arctan \frac{\frac{\pi\hbar v_F}{W}\, \big\{ { \nu+1/2 \atop \nu } \big\}
    - eV_g + \frac{e^2}{C} \sum_{\mu\neq\nu}
    \left( m_\mu - \frac{1}{2} \right) - \mu + \frac{eV}{2}}{\gamma}
  \right.\nonumber \\
&& \left.{}- \arctan \frac{\frac{\pi\hbar v_F}{W}\,\big\{ { \nu+1/2 \atop \nu
    }\big\}- eV_g + \frac{e^2}{C} \sum_{\mu\neq\nu}
    \left( m_\mu - \frac{1}{2} \right) - \mu - \frac{eV}{2}}{\gamma}
    \right)
\label{4.Rin}
\ea
for $m_\nu=0$ and $m'_\nu=1$ and
\ba
R_{\vec m\to\vec m'} & = & R_\nu^\mathrm{out}
\; := \; R_0 \left( \int_{\mu-eV/2}^\infty
  + \int_{\mu+eV/2}^\infty \right)  \frac{dE}{2\pi} \,
  \frac{2\gamma}{\big[E-E_\nu - \frac{e^2}{C} \sum_{\mu\neq\nu}
  \left( m_\mu - \frac{1}{2} \right)\big]^2 + \gamma^2} \nonumber \\
& = & \frac{R_0}{\pi} \left( \pi
  + \arctan \frac{\frac{\pi\hbar v_F}{W}\, \big\{ { \nu+1/2 \atop \nu } \big\}
    - eV_g + \frac{e^2}{C} \sum_{\mu\neq\nu}
    \left( m_\mu - \frac{1}{2} \right) - \mu + \frac{eV}{2}}{\gamma}
  \right. \nonumber \\
&& \left.{}+ \arctan \frac{\frac{\pi\hbar v_F}{W}\,\big\{ { \nu+1/2 \atop \nu
    }\big\} - eV_g + \frac{e^2}{C} \sum_{\mu\neq\nu}
    \left( m_\mu - \frac{1}{2} \right) - \mu - \frac{eV}{2}}{\gamma}
    \right)
\label{4.Rout}
\ea
\end{widetext}
for $m_\nu=1$ and $m'_\nu=0$, respectively. The notation signifies that the
integrals share the same integrand. The rate for tunneling into (out of) state
$|\nu\rangle$, $R_\nu^\mathrm{in}$ ($R_\nu^\mathrm{out}$), is proportional to
the spectral weight of this state below (above) the electrochemical potentials
in the leads. $R_0$ is a characteristic rate further discussed below. In
the present approximation, $R_0$ drops out of
the solution for the stationary state. The Pauli master equation now reads
\be
\frac{d}{dt}\, P(\vec m) = \sum_{\vec m'} \big[
  R_{\vec m'\to \vec m}\, P(\vec m') - R_{\vec m\to \vec m'}\, P(\vec m)
  \big] .
\ee
For the stationary state the left-hand side vanishes. For practical
calculations, the Fock space of the dot is truncated
by restricting the single-particle basis to
$N_\mathrm{state}$ states. We have checked that increasing this cutoff does not
lead to significant changes in the numerical results.

Like in the non-interacting case, the chemical potential and the gate voltage
only appear as $eV_g+\mu$ in Eqs.\ (\ref{4.Ifinal}), (\ref{4.Rin}), and
(\ref{4.Rout}) so that the current only depends on this combination and we can
set $\mu=0$ without loss of generality. Moreover, the current
remains an odd function of the bias voltage and an even function of the gate
voltage (for $\mu=0$). The current remains periodic in $eV_g$ but the period is
changed to $E_0 + e^2/C$, since a shift of $eV_g$ by $E_0 + e^2/C$ in Eqs.\
(\ref{4.Ifinal}), (\ref{4.Rin}), and (\ref{4.Rout}) can be
compensated for by shifting the ladder of single-particle states by one level
spacing $E_0$, taking into account that this will also increase or decrease the
excess charge in the stationary state by one unit. Moreover, the change from
parallel to antiparallel exchange fields is equivalent to a shift of $eV_g$ by
half that period.
For antiparallel exchange fields and $\mu=eV_g=0$, a conductance peak remains
pinned at zero bias, due to the particle-hole symmetry of the interacting model.
On the other hand, there is no reason for the differential conductance to remain
periodic in the \emph{bias} voltage.

Our approach is not only valid for weak coupling to the leads,
by construction, but also becomes exact in the limit of weak interactions, for
any coupling,\cite{MWL91} since then the peaks in the spectral function merge.
Moreoever, the limit of strong coupling to the leads ($\eta\to 0$) is also
correct: The broadening of the peaks in the spectral function
becomes $\gamma\gg eV$ so that the density of states in the dot becomes constant
and equal to $1/E_0 = W/\pi\hbar v_F$. Thus the current becomes $I = (e^2/h)
2\pi (\Gamma W/\pi v_F) V = (e^2/h)\, V$, the result for an open channel. Note
that this holds for arbitrary interacting strength.

\begin{figure}
\includegraphics[width=3.0in,clip]{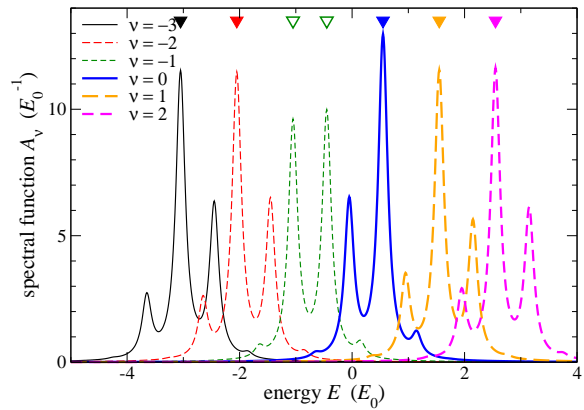}
\caption{(Color online) Spectral function $A_\nu(E)$ for several states
$|\nu\rangle$ with single-particle energies close to the Weyl node. The
parameters are $eV=2E_0$, $e^2/C = 0.6\, E_0$, and $eV_g=0.25\, E_0$, chosen so
that there are no accidental degeneracies. The
exchange fields in the barriers are parallel. The level spacing $E_0 = \pi\hbar
v_F/W$ is used as the energy unit. The energies of the strongest peaks are
marked by triangles.}
\label{fig.Aspectral}
\end{figure}

Figure \ref{fig.Aspectral} shows the spectral functions $A_\nu(E)$ for various
$\nu$. The spectral functions for
$\nu=-3,-2,1,2$, which correspond to large single-particle energies $|E_\nu|$,
have similar shapes and mainly differ in shifts in energy. Their three-peak
structure shows that three charge states of the dot dominate. Their shape is
similar since the charge fluctuations mostly take place in other states, namely
in $\nu=-1,0$. The spectral functions for $\nu=-1,0$ have only two dominant
peaks because the charge in the other states assumes mainly two distinct values;
the fluctuating occupation of the state $|\nu\rangle$ itself does not matter,
see Eq.\ (\ref{4.Anu}). Also note the phase change of the pattern for
increasing $\nu$: The spectral functions $A_\nu(E)$ for large $\nu$ have an
additional energy shift of $e^2/C$ compared to those for small $\nu$ since for
large $\nu$ the total occupation of the other states is larger by one than for
small $\nu$.

\begin{figure}
\raisebox{2ex}{(a)}\includegraphics[width=2.6in]{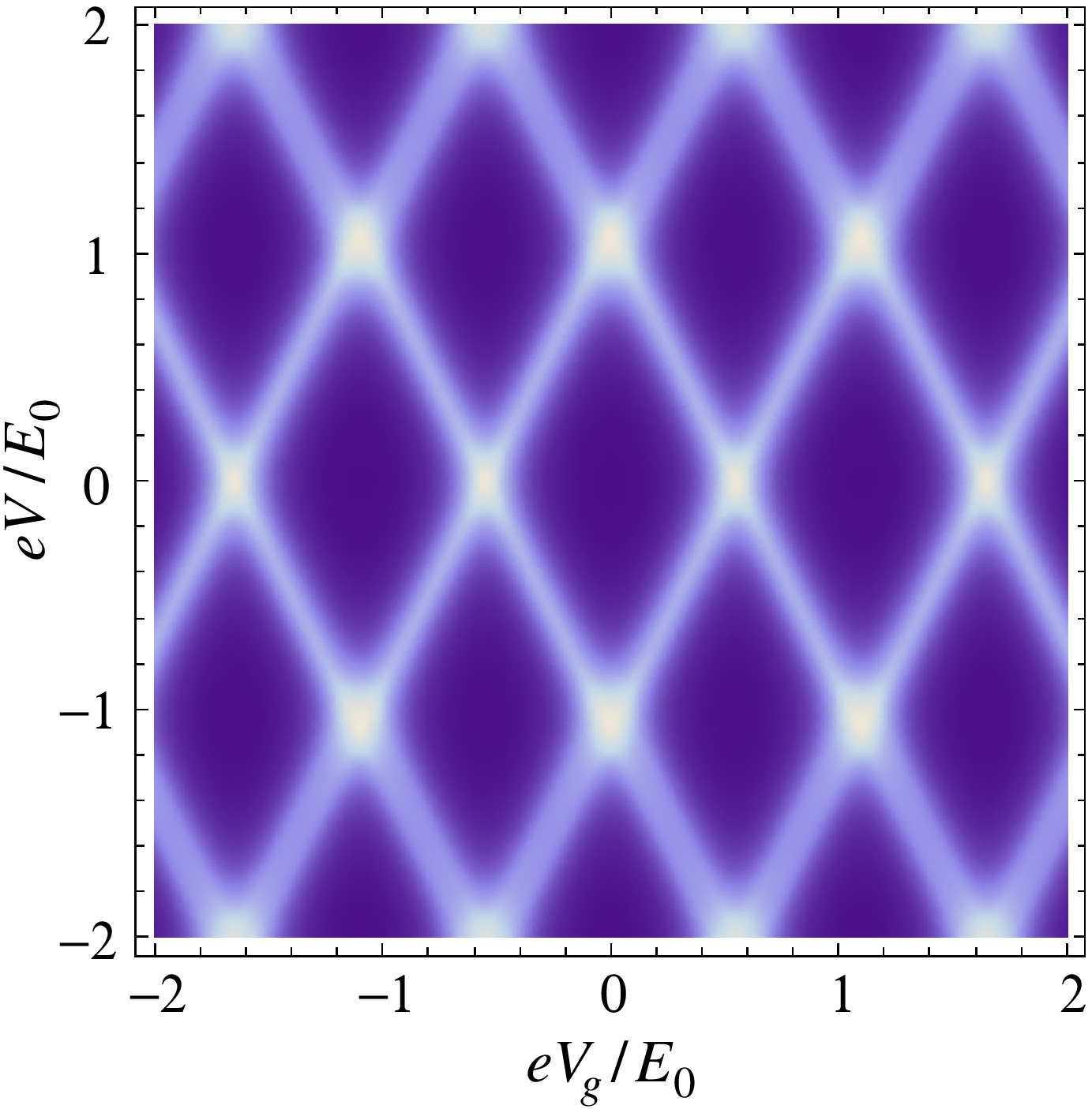}
\raisebox{2ex}{(b)}\includegraphics[width=2.6in]{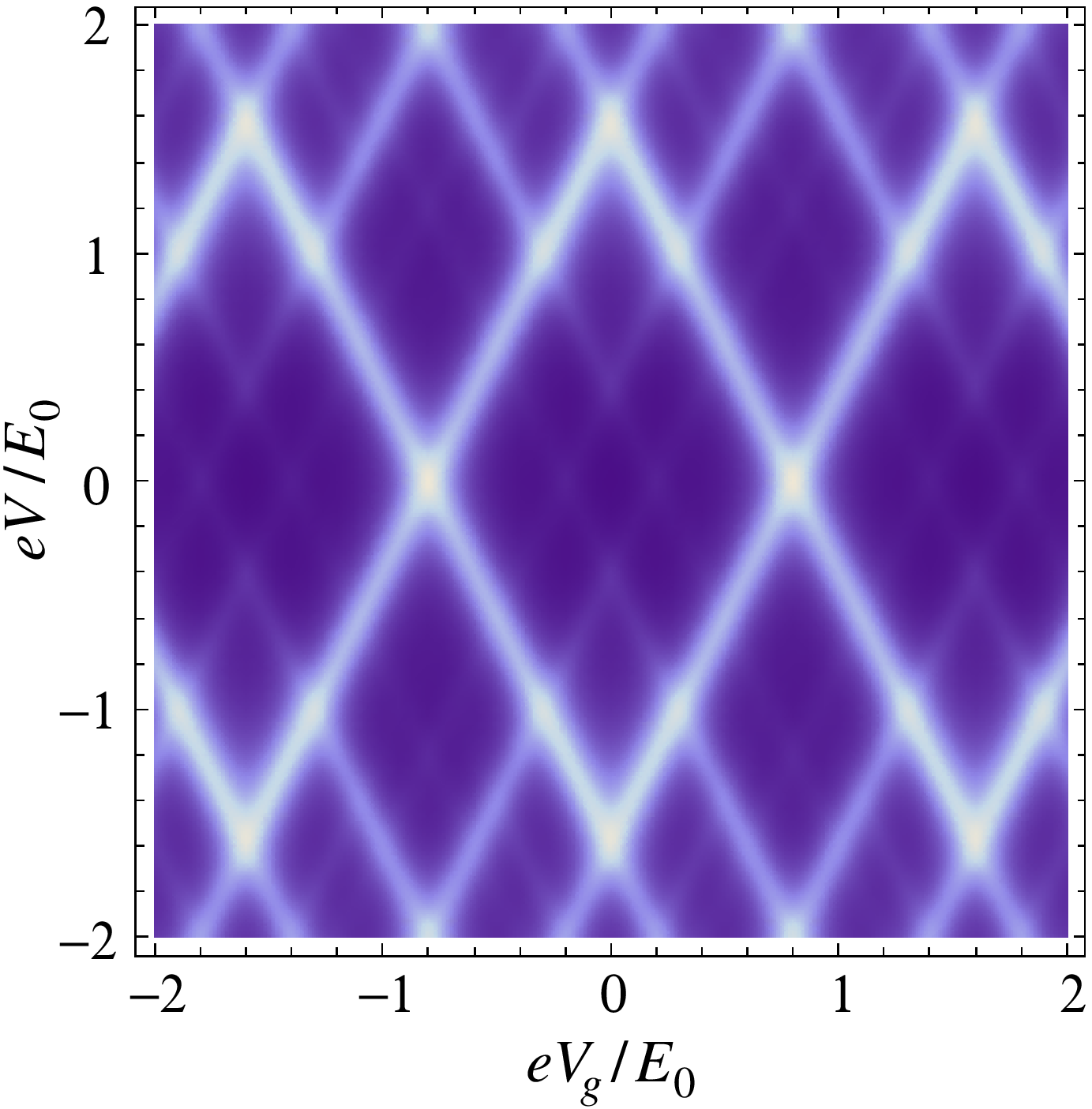}
\raisebox{2ex}{(c)}\includegraphics[width=2.6in]{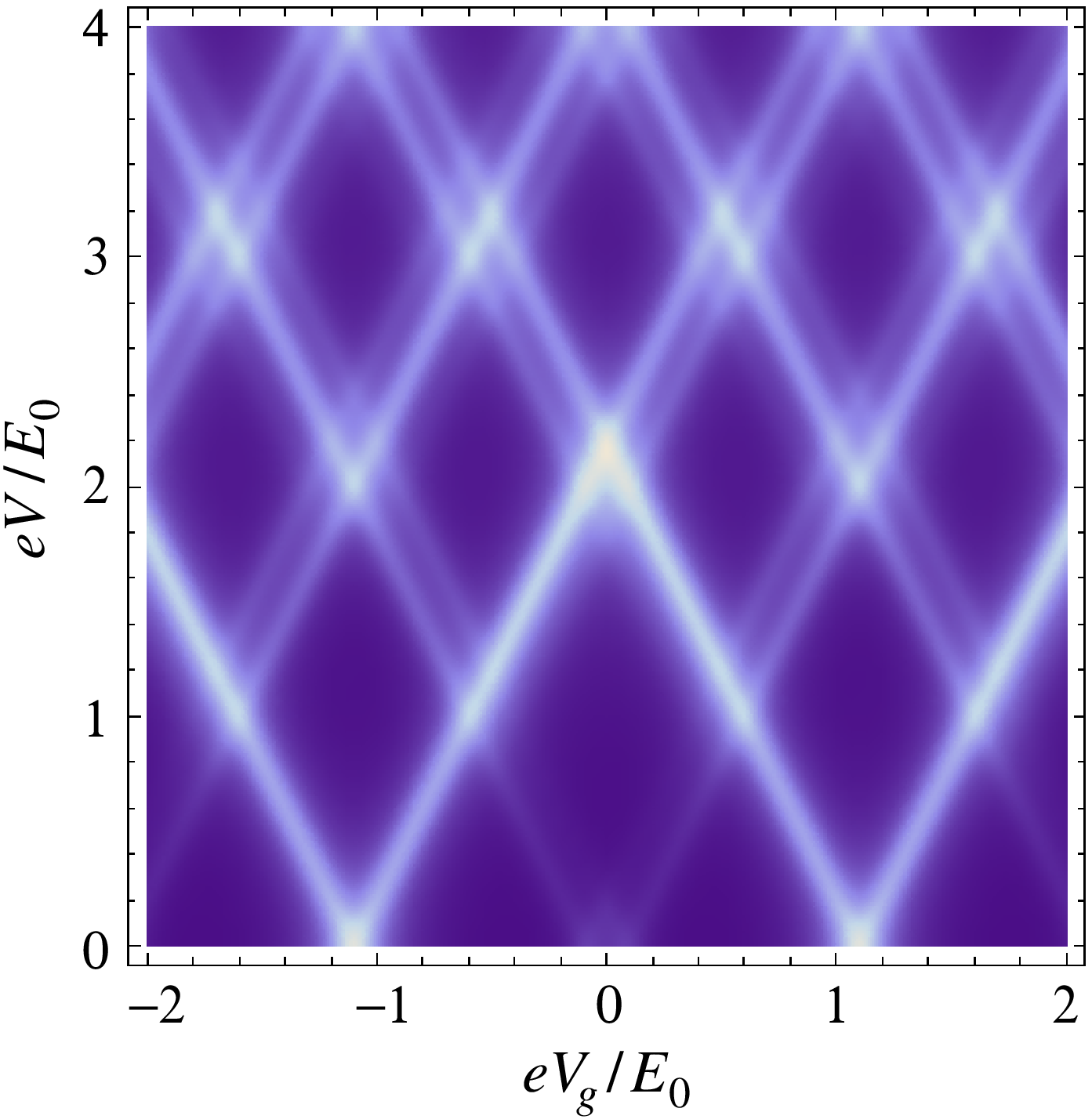}
\caption{(Color online) Density plots of the differential conductance
$dI/dV$ through a QSH quantum dot with parallel exchange fields as a function of
gate voltage $V_g$ and bias voltage $V$. The dimensionless strength of the
magnetic barriers is $\eta=1$ ($\gamma = 0.086689\,E_0$). The interaction
strength is (a) $e^2/C = 0.1\,E_0$, (b) $e^2/C = 0.6\, E_0$ (same as in Fig.\
\ref{fig.Aspectral}), and (c) $e^2/C = 1.2\, E_0$ (note the shifted vertical
axis).}
\label{fig.inter}
\end{figure}

Figure \ref{fig.inter} shows the differential conductance for various
interaction strengths and otherwise the same parameters as in Fig.\
\ref{fig.noninter}(a). With increasing interaction strength the
Pauli-blockade diamonds morph into Coulomb-blockade diamonds and become larger,
in agreement with the gate-voltage period $E_0 + e^2/C$. Also, the
sequential-tunneling peaks defining the Coulomb-blockade threshold gain more
weight; recall that in the non-interacting case $dI/dV$ was periodic in the
bias voltage $V$.

The differential-conductance plot in Fig.\ \ref{fig.inter}(b) corresponds to
the spectral functions $A_\nu(E)$ shown in Fig.\ \ref{fig.Aspectral}. The plots
are not easy to compare, since the spectral functions are calculated
for fixed bias voltage $V$, but one can say that the splitting of the peak in
$A_\nu(E)$ due to interactions is reflected by a splitting of the
sequential-tunneling peaks in $dI/dV$. Importantly, weaker conductance peaks are
visible inside the Coulomb-blockade diamonds.
While a single dot charge dominates in this regime, other dot charges have
non-zero probabilities, which are controlled by the tails of the Lorentzian
peaks. These subdominant dot charges lead to satellite peaks in the
differential conductance.

\begin{figure}
\includegraphics[width=2.6in]{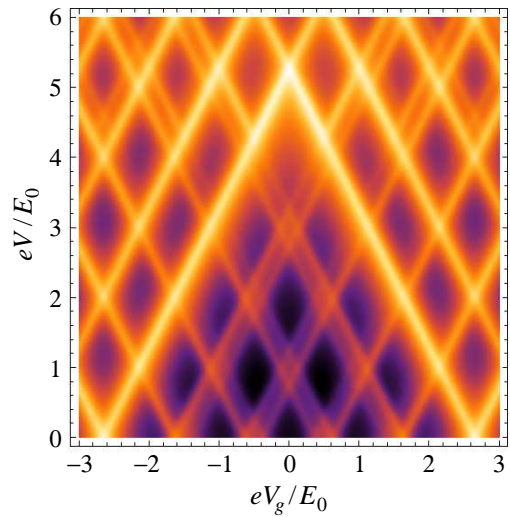}
\caption{(Color online) Density plot of the differential conductance
$dI/dV$ through a QSH quantum dot with parallel exchange fields as a function of
gate voltage $V_g$ and bias voltage $V$ for strong
electron-electron interaction $e^2/C = 4.3\,E_0$. The dimensionless strength of
the magnetic barriers is $\eta=1$.
Other than in the previous figures the logarithm of $dI/dV$ is plotted,
which enhances the contrast at small $dI/dV$. Also note the extended voltage
scales.}
\label{fig.strong}
\end{figure}

\begin{figure}
\raisebox{2ex}{(a)}\includegraphics[width=2.0in]{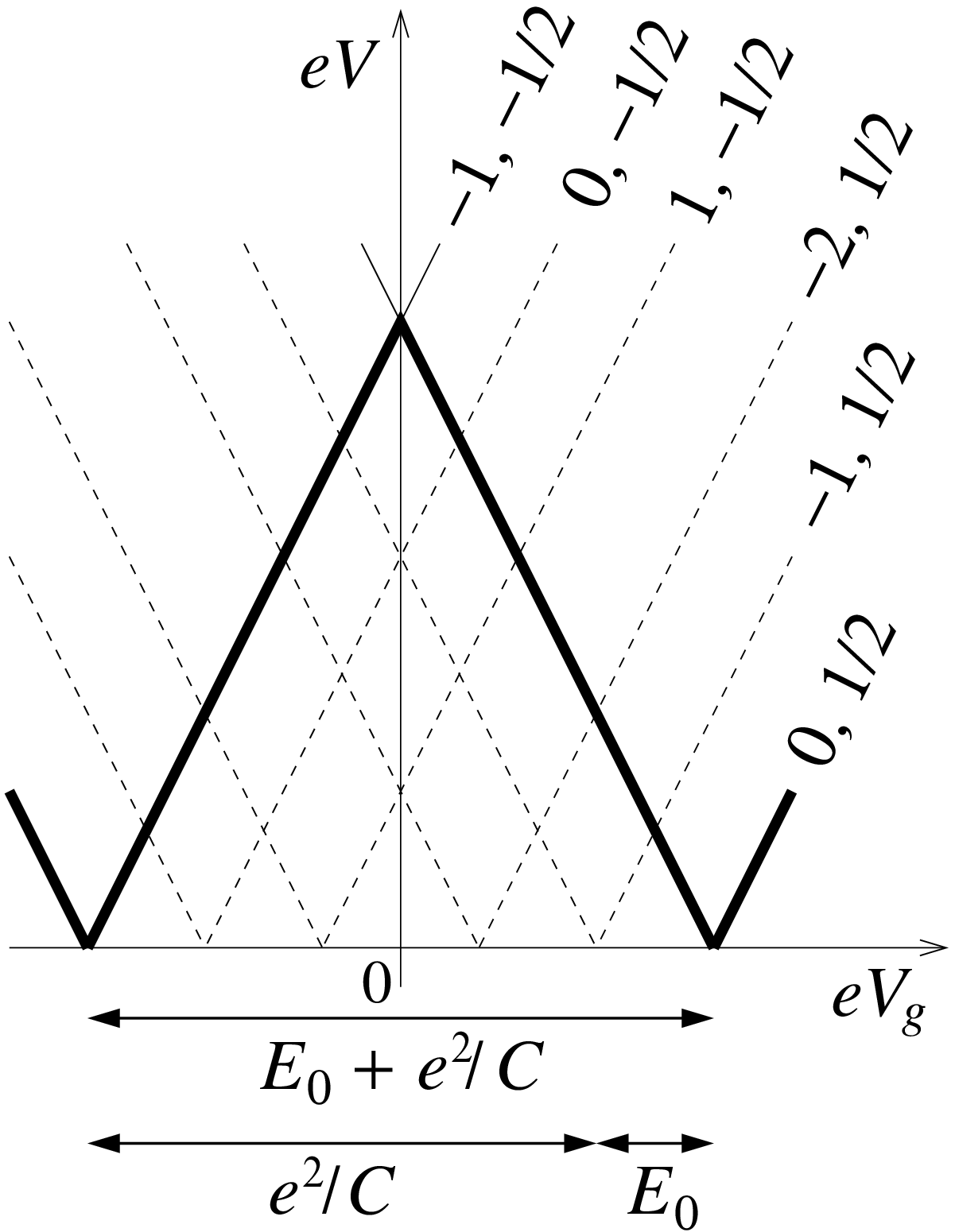}\\[2ex]
\raisebox{1ex}{(b)}\includegraphics[width=3.1in]{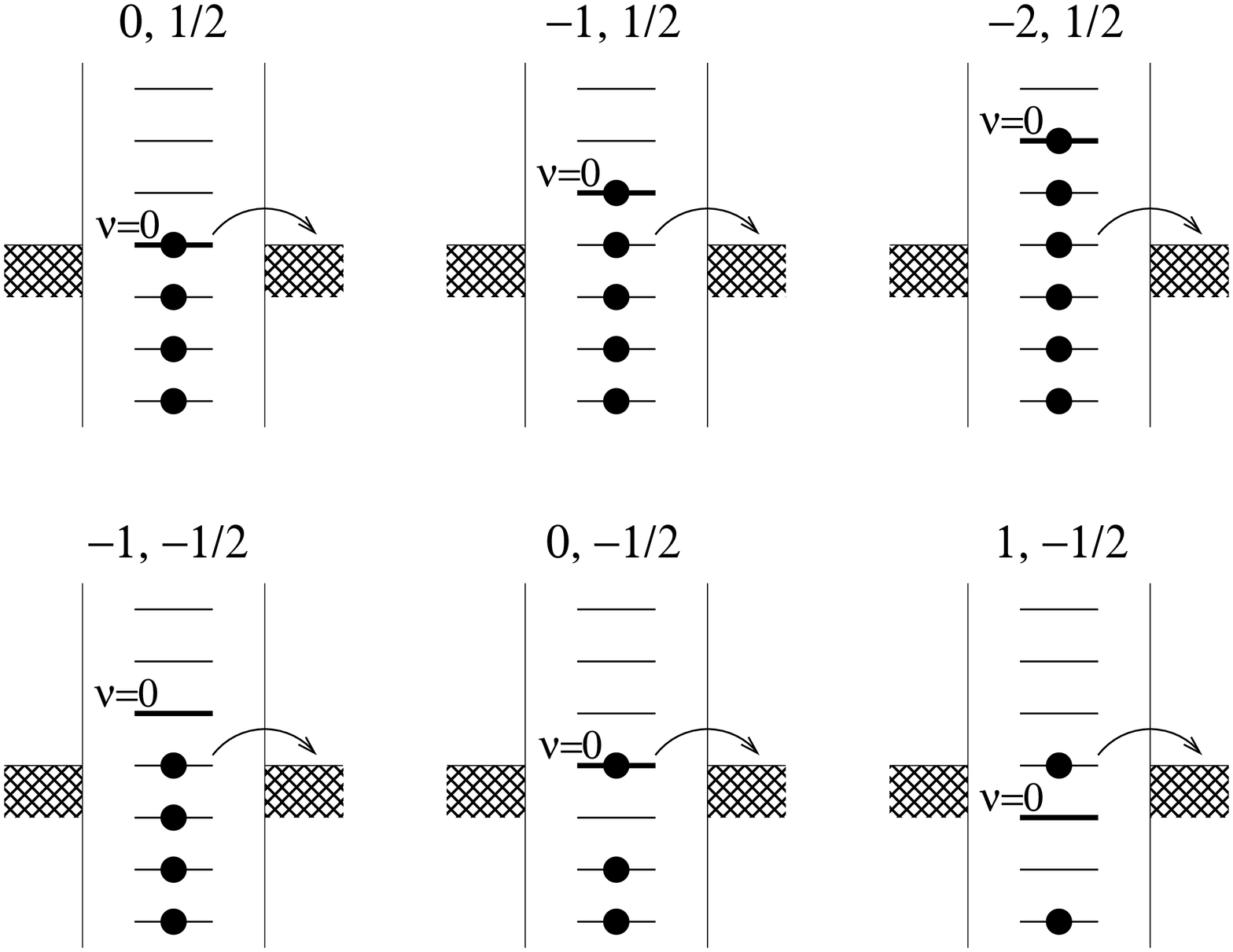}
\caption{(a) Sketch of strong sequential-tunneling lines for the case of
strong interaction, $e^2/C>E_0$. The lines sloping upward correspond to
electrons tunneling out of the quantum dot. They are labeled by the number
$\nu$ of the single-particle state involved and by the excess electron number
$\Delta n_{\neg\nu} = \sum_{\mu\neq\nu} (m_\mu-1/2)$ of the
other states. The bold lines denote the edges of the Coulomb-blockade diamonds.
(b) Illustration of the tunneling processes responsible for the lines marked in
panel (a) for $eV=0$ and $eV_g$ tuned so that the process is resonant.}
\label{fig.lines_strong}
\end{figure}

Figure \ref{fig.strong} shows $dI/dV$ for strong interaction, $e^2/C=4.3\,
E_0$. Interestingly, the conductance peaks inside the Coulomb diamond show an
approximate periodicity with the \emph{non-interacting}
periods $E_0$ in $eV_g$ and $2E_0$ in $eV$. Thus the
non-interacting pattern re-emerges for strong interactions. Further
inspection shows that there are two sets of lines, each with period $E_0$. The
spectral-function peaks and tunneling processes corresponding to these features
are illustrated in Fig.\ \ref{fig.lines_strong}. The sequential tunneling lines
are characterized by the state $|\nu\rangle$ involved in the
tunneling and by the excess occupation number $\Delta n_{\neg\nu}$ of the other
states. For strong interactions, at least one and typically only one peak
resulting from $A_\nu(E)$ for fixed $\nu$ lies inside the Coulomb-blockade
diamond. Figure \ref{fig.lines_strong} shows that there are two families of
lines corresponding to $\Delta n_{\neg\nu} = 1/2$ and $\Delta n_{\neg\nu} =
-1/2$, respectively. (Additional families exist but correspond to initial states
with larger excess charge and much smaller probability.)

\begin{figure}
\includegraphics[width=3.0in,clip]{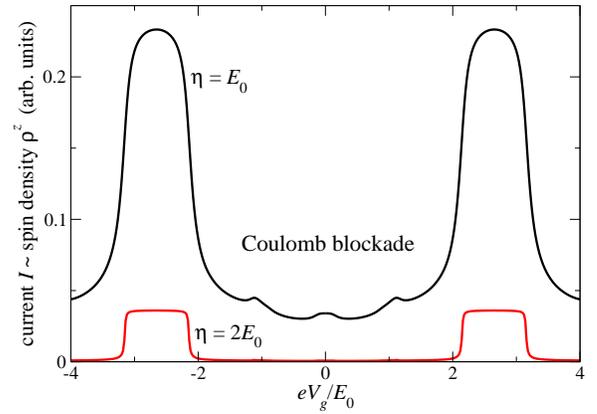}
\caption{(Color online) Current as a function of gate voltage for bias voltage
$eV=E_0$, interaction strength $e^2/C=4.3\,E_0$, and two strengths $\eta$ of the
magnetic barriers, which have parallel exchange fields. The
current is proportional to the uniform \textit{z}-component of the spin
density.}
\label{fig.curvg}
\end{figure}

We now turn to the consequences of spin-mo\-men\-tum locking.
From the continuity equation for the charge, $\partial_t\rho+\partial_x j=0$,
one easily obtains the charge current in terms of the many-particle wave
function $\Psi$,
\ba
\lefteqn{ j(x) = -e v_F \int dx_1 dx_2\ldots
\sum_{\sigma_1\sigma_2\ldots=\pm 1} \sum_i \delta(x-x_i) } \nonumber \\
&& {}\times \Psi^\ast_{\sigma_1\sigma_2\ldots}(x_1,x_2,\ldots)\,
  \sigma_i\,
  \Psi_{\sigma_1\sigma_2\ldots}(x_1,x_2,\ldots) .\quad
\ea
This expression is clearly proportional to the \textit{z}-com\-po\-nent of the
spin density $\rho^z$,
\be
j(x) = -\frac{2ev_F}{\hbar}\, \rho^z(x) .
\label{4.jrhoz}
\ee
In the stationary state, the current $j(x)=I$ is uniform so that the
\textit{z}-component of the spin density is also uniform and given by
$\rho^z = -{\hbar I}/{2ev_F}$.
Thus the quantum-dot structure permits to control the magnetization of the
whole edge with a gate voltage applied locally to the dot. We do not address
the \emph{dynamics} in this paper, i.e., we do not study how the edge reacts to
a non-adiabatic change of the gate voltage. Figure \ref{fig.curvg} shows the
uniform current (spin-\textit{z} density) as a function of the gate voltage
for strong interaction,
$e^2/C = 4.3\,E_0$, corresponding to Fig.\ \ref{fig.strong}. For stronger
magnetic barriers, the current and the spin density are reduced but the contrast
between sequential-tunneling (``on'') and Coulomb-blockade (``off'') regimes is
higher.

Equation (\ref{4.jrhoz}) begs the question whether the \textit{z}-com\-po\-nent
of the \emph{spin} current is related to the \emph{charge} density. Indeed, from
the spin continuity equation $\partial_t\rho^z + \partial_x j^z=0$ one obtains
the spin current
\ba
j^z(x) & = & \frac{\hbar v_F}{2} \int dx_1 dx_2\ldots
  \sum_{\sigma_1\sigma_2\ldots=\pm  1} \sum_i \delta(x-x_i)\, \nonumber \\
&& {}\times \Psi^\ast_{\sigma_1\sigma_2\ldots}(x_1,x_2,\ldots)\,
  \Psi_{\sigma_1\sigma_2\ldots}(x_1,x_2,\ldots) \nonumber \\
& = & -\frac{\hbar v_F}{2e}\, \rho(x) .
\ea
This result does not hold in the magnetic barriers, since spin is not conserved
there. For the same reason, the spin-\textit{z} current and thus the charge
density are generally not continuous through the barriers. An exact result can
be obtained at $eV_g=0$ for both parallel and antiparallel exchange fields: Here
we have $\rho(x)=0$ and thus $j^z(x)=0$ within the dot for arbitrary bias
voltage due to particle-hole symmetry.

The approximations involved in describing the coupling to the leads by a
non-hermitian Hamiltonian and in truncating the equations of motion for the
Green functions as shown in Appendix \ref{app.eom} may appear to be
unsystematic. Insight into their meaning can be gained from an alternative
derivation of the same final expression for the current. Since we calculate the
stationary probabilities $P(\vec m)$ in an effective tunneling
picture with bare tunneling rate $R_0$ for each lead, it is natural to write
down the current in the same picture. In the genealized sequential-tunneling
approximation including level broadening, the current through lead $\alpha$
is\cite{BrF04,KOO04,TiE06}
\be
I^\alpha = - \alpha\,\frac{e}{\hbar} \sum_{\vec m,\vec m'}
  \bigg( \sum_\nu m_\nu' - \sum_\nu m_\nu \bigg)\,
  R^\alpha_{\vec m\to \vec m'}\, P(\vec m) ,
\ee
where $\alpha=L=1$ ($\alpha=R=-1$) for the left (right) lead. $R^\alpha_{\vec
m\to \vec m'}$ denotes the part of the rates in Eqs.\ (\ref{4.Rin}) and
(\ref{4.Rout}) due to lead $\alpha$. For the symmetrized current we obtain,
using an obvious notation,
\ba
I & = & \frac{I^L + I^R}{2} \; = \; -\frac{e}{2\hbar} \sum_{\vec m} \sum_\nu
  \nonumber \\
&& {}\times \left\{\begin{array}{cl}
  R_\nu^{L,\mathrm{in}} - R_\nu^{R,\mathrm{in}} & \mbox{for $m_\nu=0$} \\[1ex]
  -R_\nu^{L,\mathrm{out}} + R_\nu^{R,\mathrm{out}} &
  \mbox{for $m_\nu=1$}
  \end{array}\right\}\, P(\vec m) .\qquad
\ea
We reobtain Eq.\ (\ref{4.Ifinal}) for the current if we assume that the scale of
the sequential-tunneling rates is the coupling function $R_0=\Gamma^L=\Gamma^R$.
This is plausible but was not \textit{a-priori} obvious since our
Hamiltonian is not of tunneling form. The approximations used
in the first approach to find the mapping of the probabilites $P(\vec m)$ to
the current are thus equivalent to a sequential-tunneling approximation with
additional level broadening. The first approach has the advantages of also
providing the spectral functions $A_\nu(E)$ and of being suitable for
approximations other than perturbation theory for weak coupling.

\section{Summary}
\label{sec.sum}

We have considered a quantum dot in a QSH edge realized by two thin magnetic
tunneling barriers with parallel or antiparallel exchange fields. For vanishing
electron-electron interactions, the linear dispersion of edge
states leads to an equidistant ladder of dot resonances and to a double
periodicity of the differential conductance in both gate and bias voltage. The
results can be analyzed within the framework of Meir-Wingreen theory
for arbitrary coupling between dot and leads although the Hamiltonian is not of
tunneling form.

In the interacting case, an equation-of-motion approach for the local retarded
Green function yields a multiple-peak structure in the spectral function, where
the peaks result from distinct charge states of the dot. Their probabilities
are calculated within a rate-equation approach, which is valid for high
magnetic tunneling barriers. The approach also becomes exact in the two limits
of vanishing tunneling barriers and of weak interactions. We have calculated and
analyzed the differential conductance from weak to strong Coulomb
interaction on the dot. For increasing interaction strength, the
Coulomb-blockade diamonds grow in size. The periodicity in the gate voltage is
retained, while the periodicity in the bias voltage is destroyed. For stronger
interactions, weak tunneling peaks appear inside the Coulomb-blockade
diamonds. For very stong interactions compared to the spacing of the dot
resonances, two copies of the \emph{non-interacting} periodic pattern re-emerge
inside the diamonds.

Spin-momentum locking leads to an obvious proportionality between the charge
current and the \textit{z}-component of the spin density or magnetization in
the whole QSH edge. This should make it possible to control the magnetization
by a locally applied gate voltage.

\acknowledgments

The author would like to thank T. Ludwig, A. P. Schnyder, E. M. Hankiewicz,
and A. Croy for useful
discussions. Financial support by the Deutsche Forschungsgemeinschaft, in part
through Research Unit 1154, \textit{Towards Molecular Spintronics}, is
acknowledged.

\appendix

\section{Evaluation of the equation of motion}
\label{app.eom}

In the following solution of the equation of motion (\ref{4.EOM}) we set
$\hbar=1$. Equation (\ref{4.EOM}) contains the commutator
\begin{widetext}
\be
[H_\mathrm{int},c_\nu] = \frac{e^2}{2C} \sum_{\mu\mu'}
  \left[ \left(n_{\mu}-\frac{1}{2}\right) \left(n_{\mu'}-\frac{1}{2}\right),
  c_\nu \right]
  = - \frac{e^2}{C} \sum_{\mu\neq \nu} \left( n_{\mu} - \frac{1}{2} \right)\,
  c_\nu .
\ee
Noting that $n_{\mu}$ and $H_\mathrm{int}$ commute with $K_\mathrm{dot}$, it
follows that
\ba
e^{iK_\mathrm{dot}t} \, [H_\mathrm{int},c_\nu]\, e^{-iK_\mathrm{dot}t}
& = & - \frac{e^2}{C} \sum_{\mu\neq \nu} \left( n_{\mu} - \frac{1}{2} \right)\,
  e^{iK_\mathrm{dot}t} \, c_\nu \, e^{-iK_\mathrm{dot}t} \nonumber \\
& = & - \frac{e^2}{C}\, e^{-i E_\nu t -\gamma t} \sum_{\mu\neq \nu}
  \left( n_{\mu} - \frac{1}{2} \right)\,
  e^{iH_\mathrm{int}t} c_\nu e^{-iH_\mathrm{int}t} \nonumber \\
& = & - \frac{e^2}{C}\, e^{-i E_\nu t -\gamma t} \sum_{\mu\neq \nu}
  \left( n_{\mu} - \frac{1}{2} \right)\,
  \exp\Bigg( -i\, \frac{e^2}{C} \sum_{\mu'\neq \nu}
  \left( n_{\mu'} - \frac{1}{2} \right) t \Bigg)\, c_\nu .
\ea
Next, we have to evaluate the anticommutator
$\big\{ e^{iK_\mathrm{dot}t}\,
  [H_\mathrm{int}, c_\nu] \, e^{-iK_\mathrm{dot}t} ,
  c_{\nu'}^\dagger \big\}$.
For the diagonal case $\nu=\nu'$, we immediately get
\ba
\lefteqn{ \{ e^{iK_\mathrm{dot}t}\,
  [H_\mathrm{int}, c_\nu] \, e^{-iK_\mathrm{dot}t} ,
  c_\nu^\dagger \}
= - \frac{e^2}{C}\, e^{-i E_\nu t -\gamma t} \sum_{\mu\neq \nu}
  \left( n_{\mu} - \frac{1}{2} \right)\,
  \exp\Bigg( -i\, \frac{e^2}{C} \sum_{\mu'\neq \nu}
  \left( n_{\mu'} - \frac{1}{2} \right) t \Bigg) } \nonumber \\
&& = - \frac{e^2}{C}\, e^{-i E_\nu t -\gamma t} \sum_{\mu\neq \nu}
  \left( n_{\mu} - \frac{1}{2} \right)\,
  \prod_{\mu'\neq \nu} \left[
  \exp\left(i\,\frac{e^2}{2C}\, t\right)\, (1-n_{\mu'})
  + \exp\left(-i\,\frac{e^2}{2C}\, t\right)\, n_{\mu'} \right] \nonumber \\
&& = -i\, e^{-i E_\nu t -\gamma t}\, \frac{\partial}{\partial t}\,
  \prod_{\mu\neq \nu} \left[
  \exp\left(i\,\frac{e^2}{2C}\, t\right)\, (1-n_\mu)
  + \exp\left(-i\,\frac{e^2}{2C}\, t\right)\, n_\mu \right] ,\hspace{10em}
\ea
while for $\nu\neq\nu'$ we find
\ba
\lefteqn{ \{ e^{iK_\mathrm{dot}t}\,
  [H_\mathrm{int}, c_\nu] \, e^{-iK_\mathrm{dot}t} ,
  c_{\nu'}^\dagger \}
= \frac{e^2}{C}\, e^{-iE_\nu t -\gamma t} \, \Bigg[
  \sum_{\mu\neq \nu} \left( n_\mu - \frac{1}{2}\right)\,
  \exp\Bigg( - i\, \frac{e^2}{C} \sum_{\mu'\neq\nu}
  \left( n_{\mu'} - \frac{1}{2} \right) t \Bigg),
  c_{\nu'}^\dagger c_\nu \Bigg] } \nonumber \\
&& = 2\, e^{-iE_\nu t -\gamma t} \, \frac{\partial}{\partial t}\,
  \sin\left(\frac{e^2}{2C}\, t\right)\,
  \prod_{\mu\neq \nu,\nu'} \left[
  \exp\left(i\,\frac{e^2}{2C}\, t\right)\, (1-n_\mu)
  + \exp\left(-i\,\frac{e^2}{2C}\, t\right)\, n_\mu \right] \,
  c_{\nu'}^\dagger c_\nu .\hspace{5em}
\label{A.antioff}
\ea
The equation of motion for the diagonal components of the Green function thus
reads
\ba
\lefteqn{ (E-E_\nu+i\,\gamma)\, G^\mathrm{ret}_{\nu\nu}(E) = 1
  + \int_0^\infty dt\, e^{-i E_\nu t -\gamma t}\, \frac{\partial}{\partial t}\,
  \left\langle
  \prod_{\mu\neq \nu} \left[
  \exp\left(i\,\frac{e^2}{2C}\, t\right)\, (1-n_\mu)
  + \exp\left(-i\,\frac{e^2}{2C}\, t\right)\, n_\mu \right]
  \right\rangle } \nonumber \\
&& = - \int_0^\infty dt\, i\,(E-E_\nu+i\gamma)\, e^{i (E-E_\nu) t - \gamma t} \,
  \left\langle
  \prod_{\mu\neq \nu} \left[
  \exp\left(i\,\frac{e^2}{2C}\,t\right)\, (1-n_\mu)
  + \exp\left(-i\, \frac{e^2}{2C}\,t\right)\, n_\mu \right]
  \right\rangle .\hspace{3em}
\ea
Hence, the diagonal Green function is
\ba
G^\mathrm{ret}_{\nu\nu}(E) & = & -i \int_0^\infty dt\,
  e^{i (E-E_\nu) t - \gamma t} \,
  \left\langle
  \prod_{\mu\neq \nu} \left[
  \exp\left(i\,\frac{e^2}{2C}\,t\right)\, (1-n_\mu)
  + \exp\left(-i\, \frac{e^2}{2C}\,t\right)\, n_\mu \right]
  \right\rangle \nonumber \\
& = & -i \int_0^\infty dt\, e^{i (E-E_\nu) t - \gamma t}
  \sum_{\ldots,m_0,m_1,\ldots=0}^1
  \left\langle
  \prod_{\mu\neq \nu} \exp\left((-1)^{m_\mu} i \,\frac{e^2}{2C}\, t\right)\,
  \left\{\begin{array}{cl}
    1-n_\mu & \mbox{for $m_\mu=0$} \\
    n_\mu   & \mbox{for $m_\mu=1$}
  \end{array}\right\} \right\rangle \nonumber \\
& = & \sum_{\ldots,m_0,m_1,\ldots=0}^1
  \left\langle \prod_{\mu\neq \nu}
  \left\{\begin{array}{cl}
    1-n_\mu & \mbox{for $m_\mu=0$} \\
    n_\mu   & \mbox{for $m_\mu=1$}
  \end{array}\right\} \right\rangle\,
  \frac{1}{E-E_\nu - \frac{e^2}{C} \sum_{\mu\neq\nu}
  \left( n_\mu - \frac{1}{2} \right) + i\gamma} .\hspace{5em}
\ea
The sum is over all $m_\mu$ with $\mu\neq \nu$.

For the off-diagonal components, $\nu\neq\nu'$, inserting Eq.\
(\ref{A.antioff}) into the equation of motion (\ref{4.EOM}) gives
\ba
\lefteqn{ (E-E_\nu+i\,\gamma)\, G^\mathrm{ret}_{\nu\nu'}(E) =
  2i \int_0^\infty dt\, e^{-i E_\nu t -\gamma t}\, \frac{\partial}{\partial t}\,
  \sin\left(\frac{e^2}{2C}\, t\right) } \nonumber \\
&& \quad{}\times \left\langle
  \prod_{\mu\neq \nu,\nu'} \left[
  \exp\left(i\,\frac{e^2}{2C}\, t\right)\, (1-n_\mu)
  + \exp\left(-i\,\frac{e^2}{2C}\, t\right)\, n_\mu \right] \,
  c_{\nu'}^\dagger c_\nu
  \right\rangle \nonumber \\
&& = 2 \int_0^\infty dt\, (E-E_\nu+i\gamma)\, e^{i (E-E_\nu) t - \gamma t} \,
  \sin\left(\frac{e^2}{2C}\, t\right) \nonumber \\
&& \quad{}\times \left\langle
  \prod_{\mu\neq \nu,\nu'} \left[
  \exp\left(i\,\frac{e^2}{2C}\, t\right)\, (1-n_\mu)
  + \exp\left(-i\,\frac{e^2}{2C}\, t\right)\, n_\mu \right] \,
  c_{\nu'}^\dagger c_\nu
  \right\rangle .
\ea
Performing the integral, we obtain the off-diagonal Green function
\ba
G^\mathrm{ret}_{\nu\nu'}(E) & = & \sum_{\ldots,m_0,m_1,\ldots=0}^1
  \left\langle \prod_{\mu\neq \nu,\nu'}
  \left\{\begin{array}{cl}
    1-n_\mu & \mbox{for $m_\mu=0$} \\
    n_\mu   & \mbox{for $m_\mu=1$}
  \end{array}\right\} \, c_{\nu'}^\dagger c_\nu \right\rangle
  \nonumber \\
&& {}\times \left(
  \frac{1}{E-E_\nu - \frac{e^2}{C} \sum_{\mu\neq\nu}
  \left( n_\mu - \frac{1}{2} \right) + \frac{e^2}{2C} + i\gamma}
  - \frac{1}{E-E_\nu - \frac{e^2}{C} \sum_{\mu\neq\nu}
  \left( n_\mu - \frac{1}{2} \right) - \frac{e^2}{2C} + i\gamma}
  \right) ,
\label{A.Goff}
\ea
\end{widetext}
where the sum is now over all $m_\mu$ with $\mu\neq \nu,\nu'$. The off-diagonal
components of the Green function thus contain averages involving the
off-diagonal operator $c_{\nu'}^\dagger c_\nu$. In the stationary state such
averages vanish: For the
average in Eq.\ (\ref{A.Goff}) to be non-zero, the stationary-state
reduced density operator $\rho$ would have to be non-diagonal in the
occupation-number basis. Specifically, superpositions of states
different by a single electron being transferred from $|\nu\rangle$ to
$|\nu'\rangle$ would have to occur. From the decoupled dot Hamiltonian, such
off-diagonal components of $\rho$ obtain a rotating phase factor with frequency
$\omega=|E_{\nu'}-E_\nu|/\hbar\neq 0$. For the stationary state, this time
dependence would have to be canceled by the time dependence due to the
coupling to the leads. That this cannot happen follows for example from the
fact that the leads do not contain the energy scale $E_0 = \pi\hbar v_F/W$.


\begin{thebibliography}{99}
\bibitem{footnote.BC}Due to spin-momentum locking, the probability current
density equals $v_F \psi^\dagger \sigma^z\psi$ and is automatically continuous
when the boundary condition (\ref{2.BC1}) is satisfied.
\bibitem{footnote.HF}The spurious self-interaction term appearing in the Hartree
approximation is removed by the Fock term.
\end{thebibliography}
\end{document}